\documentclass[preprint,showpacs,preprintnumbers,amsmath,amssymb]{revtex4}

\usepackage{graphicx}% Include figure files
\usepackage{dcolumn}% Align table columns on decimal point
\usepackage{bm}% bold math

\begin{document}

\preprint{APS/123-QED}

\title{Rapidity distribution as a probe for elliptical flow at intermediate energies\\}% Force line breaks with \\

\author{Sanjeev Kumar}
\author{Varinderjit Kaur}
\author{Suneel Kumar}%
 \email{suneel.kumar@thapar.edu}
\affiliation{%
School of Physics and Materials Science, Thapar University, Patiala-147004, Punjab (India)\\
}%

\date{\today}% It is always \today, today,

\begin{abstract}
Interplay between the spectator and participant matter in heavy-ion collisions
is investigated within isospin dependent quantum molecular dynamics (IQMD) model in term of 
rapidity distribution of light charged particles. The effect of different type and size rapidity distributions  
is studied in elliptical flow. 
The elliptical flow patterns show important role of the nearby spectator matter on 
the participant zone. This role is further explained on the basis of passing time of the spectator and expansion 
time of
the participant zone. The transition from the in-plane to out-of-plane is observed only when the mid-rapidity region is 
included in the rapidity bin, otherwise no transition occurs. The transition energy is found to be highly sensitive 
towards the size of the rapidity bin, while weakly on the type of the rapidity distribution. The theoretical results
are also compared with the experimental findings and are found in good agreement.\\ 
\end{abstract}

\maketitle

\section{Introduction}

Since last many years, investigation about the nuclear equation of state (NEOS) at the extreme conditions of density and 
temperature has been one of the primary driving forces in heavy ion studies at intermediate energies. The interest in low
energies, however, are for isospin effects in fusion process \cite{Aich91,Kuma105}. 
These investigations has been performed with the help of rare phenomena such as 
multifragmentation, collective flow, particle production as well as 
nuclear stopping \cite{Kuma105,Stoc86,West935,Luka055,Andr055,Luka045,Sood065,Chen065,
Sing005,Saks105,Zhan065,Huan93,Gyul835}.   
The relation between the nuclear EOS and 
flow phenomena has been explored extensively in the simulations.

Recently the analysis of transverse-momentum dependence of elliptical flow has also been put forwarded \cite{Luka055, Andr055, Dani00}. The elliptical
flow is shaped by the interplay between the geometry and mean field and, when gated by the transverse
momentum, reveals the momentum dependence of the mean field at supra-normal densities.
The parameter of the elliptic
flow is quantified by the second-order Fourier coefficient \cite{Volo965}
\begin{equation}
v_2~=~ <cos2\phi>~=~\langle\frac{p_x^2 - p_y^2}{p_x^2 + p_y^2}\rangle,
\end{equation}
from the azimuthal distribution of
detected particles at mid rapidity as

\begin{equation}
\frac{dN}{d\phi} = p_0(1 + 2v_1cos\phi + 2v_2cos2\phi+.....),
\end{equation}

where $p_x$ and $p_y$ are the ${\it x}$ and ${\it y}$ components of  momentum. The 
$p_x$ is in the reaction plane, while, $p_y$ is 
perpendicular to the reaction plane and $\phi$ is the azimuthal angle of emitted particles momentum relative to the x-axis. The positive values of $<cos 2\phi>$
reflect preferential in-plane emission, while negative values
reflect preferential out-of-plane emission. The pulsating of
sign observed recently at intermediate  energies
has received particular attention as it reflects the increased
pressure buildup in the non isotropic collision zone \cite{Adle03}.\\ 
After the pioneering
measurements at Saturne \cite{Demo90} and Bevalac \cite{Gutb89}, a wealth of experimental results have been
obtained at Bevalac and SIS \cite{Luka055,Andr055,Luka045,Wang96} as well as at AGS \cite{Pink995}, 
SPS \cite{Adle03}
and RHIC \cite{Acke01}.  In recent years, the FOPI, INDRA, and PLASTIC
BALL Collaborations \cite{Luka055, Andr055, Luka045} are actively involved in measuring
the excitation function of elliptical flow from Fermi
energies to relativistic one. In most of these studies, collisions of 
$_{79}Au^{197}~+~_{79}Au^{197}$ is undertaken. Interestingly,
elliptical flow was reported to change from positive (in-plane) taken.
negative (out-of-plane) values around 100 MeV/nucleon and maximum squeeze-out is observed around 400 MeV/nucleon. 
These observations were reported recently by us and others theoretically \cite{Luka055, Andr055,Zhan065,Dani00}. A careful analysis reveals that elliptic flow is very sensitive towards the choice of rapidity cut which makes the difference between spectator and participant matter.\\

The elliptical flow pattern of participant matter is affected by the presence of cold
spectators \cite{Dani00}. When nucleons are decelerated in the participant
region, the longitudinal kinetic energy associated with the initial colliding nuclei is converted
into the thermal and potential compression energy. In a subsequent rapid expansion(or
explosion), the collective transverse energy develops \cite{Dani00} and many particles from
the participant region get emitted into the transverse directions. The particles emitted towards
the reaction plane can encounter the cold spectator pieces and, hence, get redirected. In
contrast, the particles emitted essentially perpendicular to the reaction plane are largely
unhindered by the spectators. Thus, for the beam energies leading to rapid expansion in the
vicinity of the spectators, elliptic flow directed out of the reaction plane (squeeze-out) is
expected. This squeeze-out is related with the pace at which expansion develops, and is,
therefore, related to the EOS. This contribution of the participant and spectator matter \cite{Dani00} in the intermediate 
energy heavy-ion collisions motivated 
us to perform a detailed analysis of the excitation function of elliptical flow over different 
regions of participant and spectator matter.
If excitation function is found to be affected by the different contributions it will 
definitely, also affect the in-plane to out-of-plane emission 
i.e. transition energy. \\ 
The rapidity distribution is an important parameter to study the participant-spectator contribution in the intermediate 
energy heavy-ion collisions \cite{Sood09}. 
In this paper, we will study the effect of participant and spectator matters in term of different rapidity distributions on the  excitation function of elliptical flow. Attempts shall also be made to parameterize the transition energy in term of different rapidity bins. \\
The entire work is carried out in the framework of isospin-dependent quantum molecular dynamics (IQMD) \cite{Hart98}. 
The IQMD model is discussed 
in detail in the following section.\\
%%%%%%%%%%%%%%%%%%%%%%%%%%%%%%%%%%%%%%%%%%%%%%%%%%%%%%%%%%%%%%%%%%%%%%%%%%%%%%%%%%%%%%%%%%%%%
\section{ISOSPIN-dependent QUANTUM MOLECULAR DYNAMICS (IQMD) MODEL}
The IQMD model \cite{Saks105,Hart98}, which is an improved version of QMD model \cite{Aich91} developed by J. Aichelin and coworkers, then has been used successfully to various phenomena such as collective flow, disappearance of flow, fragmentation \& elliptical flow      

 \cite{Stoc86,Sood065,Sood09,Acke01}. 
The isospin degree of freedom enters into the calculations via symmetry potential, cross-sections and
Coulomb interaction \cite{Saks105,Hart98}.
The details about the elastic and inelastic cross-sections
for proton-proton and neutron-neutron collisions can be found in Ref. \cite{Hart98}. \\
In IQMD model, the nucleons of target and projectile
interact via two and three-body Skyrme forces, Yukawa potential and Coulomb interactions. 
In addition to the use of explicit charge states of all baryons and mesons, a symmetry potential between 
protons and neutrons corresponding to the Bethe- Weizsacker mass formula has been included.\\
The hadrons propagate using classical Hamilton equations of motion:
\begin{equation}
\frac{d\vec{r_i}}{dt}~=~\frac{d\it{\langle~H~\rangle}}{d\vec{p_i}}~~;~~\frac{d\vec{p_i}}{dt}~=~-\frac{d\it{\langle~H~\rangle}}{d\vec{r_i}},
\end{equation}
with
\begin{eqnarray}
\langle~H~\rangle&=&\langle~T~\rangle+\langle~V~\rangle\nonumber\\
&=&\sum_{i}\frac{p_i^2}{2m_i}+
\sum_i \sum_{j > i}\int f_{i}(\vec{r},\vec{p},t)V^{\it ij}({\vec{r}^\prime,\vec{r}})\nonumber\\
& &\times f_j(\vec{r}^\prime,\vec{p}^\prime,t)d\vec{r}d\vec{r}^\prime d\vec{p}d\vec{p}^\prime .
\end{eqnarray}
 The baryon-baryon potential $V^{ij}$, in the above relation, reads as:
\begin{eqnarray}
V^{ij}(\vec{r}^\prime -\vec{r})&=&V^{ij}_{Skyrme}+V^{ij}_{Yukawa}+V^{ij}_{Coul}+V^{ij}_{sym}\nonumber\\
&=&\left(t_{1}\delta(\vec{r}^\prime -\vec{r})+t_{2}\delta(\vec{r}^\prime -\vec{r})\rho^{\gamma-1}
\left(\frac{\vec{r}^\prime +\vec{r}}{2}\right)\right)\nonumber\\
& & +~t_{3}\frac{exp(|\vec{r}^\prime-\vec{r}|/\mu)}{(|\vec{r}^\prime-\vec{r}|/\mu)}
~+~\frac{Z_{i}Z_{j}e^{2}}{|\vec{r}^\prime -\vec{r}|}\nonumber\\
& &+t_{6}\frac{1}{\varrho_0}T_3^{i}T_3^{j}\delta(\vec{r_i}^\prime -\vec{r_j}).
\label{s1}
\end{eqnarray}
Here $Z_i$ and $Z_j$ denote the charges of $i^{th}$ and $j^{th}$ baryon, and $T_3^i$, $T_3^j$ are their respective $T_3$
components (i.e. 1/2 for protons and -1/2 for neutrons).
The parameters $\mu$ and $t_1,.....,t_6$ are adjusted to the real part of the nucleonic optical potential. 
For the density
dependence of nucleon optical potential, standard Skyrme-type parameterization is employed.
The potential part resulting from the convolution of the distribution function 
with the Skyrme interactions $V_{\it Skyrme}$ reads as :
\begin{equation}
{\it V}_{Skyrme}~=~\alpha\left(\frac{\rho_{int}}{\rho_{0}}\right)+\beta\left(\frac{\rho_{int}}{\rho_{0}}\right)^{\gamma}
~~\cdot
\end{equation}
The two of the three parameters of equation of state are determined by demanding that at normal nuclear matter density,
 the binding energy should be equal to 16 MeV. The third parameter $\gamma$ is usually treated as a free parameter.
 Its value is given in term of the compressibility:
\begin{equation}
\kappa~=~9\rho^{2}\frac{\partial^{2}}{\partial\rho^{2}}\left(\frac{E}{A}\right)~~\cdot
\end{equation}
The different values of compressibility give rise to Soft and Hard equations of state. It is worth mentioning that Skyrme 
forces are very successful in analysis of low energy phenomena like fusion, fission and cluster-radioactivity. \\
As noted  \cite{Zhan02}, elliptical flow is weakly affected by the choice of equation of state. On the other hand, in
the refs. \cite{Sood065,Sood09,Mage00}, the hard equation of state is used to study the directed as well as elliptical flow. 
For the present analysis, a hard (H) equation of state
has been employed along with standard energy dependent cross-section.\\
%%%%%%%%%%%%%%%%%%%%%%%%%%%%%%%%%%%%%%%%%%%%%%%%%%%%%%%%%%%%%%%%%%%%%%%%%%%%%%%%%%%%%%%%%%%%%
\section{Results and Discussion}
\begin{figure}
\includegraphics{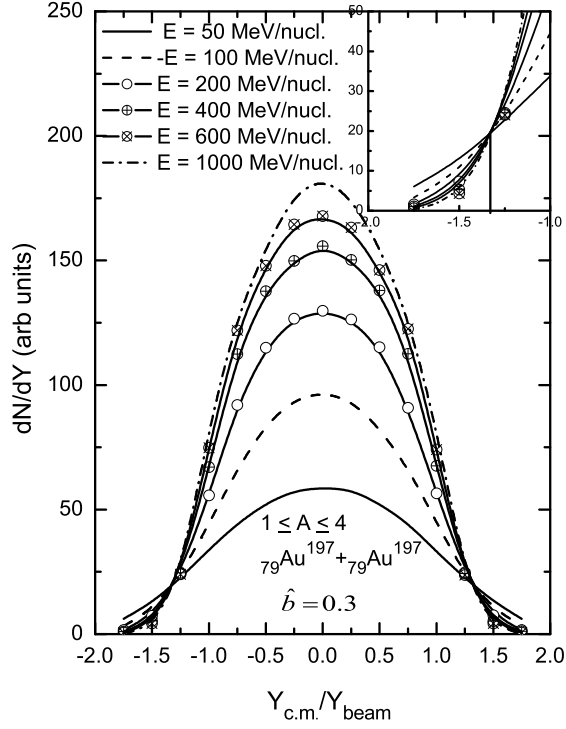}
\caption{\label{fig:1}The rapidity distribution dN/dY as a function of $Y_{c.m.}/Y_{beam}$ at different incident 
energies ranging from 50 to 1000 MeV/nucleon for light charged particles (LCP's)[1$\le$ A $\le$ 4]. The 
inset is showing only one side of the rapidity distribution.}
\end{figure}

For the present analysis, simulations are carried out for thousand of events for the reaction of 
$_{79}Au^{197}~+~_{79}Au^{197}$ at semi-central geometry using a hard equation of state. The whole 
of the analysis is performed for light charged particles (LCP's)[1$\le$ A $\le$ 4]. The reaction conditions 
and fragments are 
chosen on the basis of availability of experimental data \cite{Luka055,Andr055,Luka045}. 
As noted in Ref. \cite{Lehm93}, the relativistic effects do not play
role at these incident energies and the intensity of sub-threshold particle production is 
very small. The phase space generated by the IQMD model has been analyzed using 
the minimum spanning
tree (MST) \cite{Sood065} method. The MST method binds two nucleons in a fragment if their distance is 
less than 4 fm. This is the one of the 
widely used method in the intermediate energy heavy-ion collisions. However, 
some improvements like momentum and binding energy are also discussed in the literature \cite{Sood065}. 
In recent years, more sophisticated and complicated algorithms are also available in the literature \cite{saca}.
The entire calculations are performed at $t = 200$ fm/c. This time is chosen by keeping in view the saturation of 
the collective flow \cite{Sood065}.\\
%%%%%%%%%%%%%%%%%%%%%%%%%%%%%%%%%%%%%%%%%%%%%%%%%%%%%%%%%%%%%%%%%%%%%%%%%%%%%%%%%%%%%%%%%%%%%%%%5
As our purpose of present study is to understand the effect of participant-spectator matter on the excitation function of 
elliptical flow, one will concentrate on the rapidity distribution only \cite{Sood09}. The important concept of spectators and participants in collisions was first introduced by
Bowman et al. \cite{Bowm73} and later employed for the description of a wide-angle energetic particle
emission by Westfall et al. \cite{West76}. The two nuclei slamming against one other can be viewed as
producing cylindrical cuts through each other. The swept-out nucleons or participants (from
projectile and target) undergo a violent collision process. In the meantime, the remnants of the projectile
and target continue with largely undisturbed velocities, and
are much less affected by the collision process than the participant nucleons. On one hand,
this picture is supported by features of the data \cite{Andr055} and, on the other, by dynamic simulations \cite{Gait00}. 
During the violent stage of a reaction, the spectators can influence the behavior
of participant matter. Following the same, we are dividing the rapidity distribution in the different cuts 
in term of $Y_{c.m.}/Y_{beam}$ parameter, which is given as: 
\begin{equation}
   Y(i)=\frac{1}{2}~ln\frac{E(i)+P_z(i)}{E(i)-P_z(i)}
\end{equation}
where E(i) and $P_z(i)$ are respectively, the total energy and longitudinal momentum of $i^{th}$ particle.\\
%%%%%%%%%%%%%%%%%%%%%%%%%%%%%%%%%%%%%%%%%%%%%%%%%%%%%%%%%%%%%%%%%%%%%%%%%%%%%%%%%%%
As we are interested in studying the effect of rapidity distribution on the incident energy dependence of elliptical 
flow, one has to understand the 
$dN/dY$ as a function of $Y_{c.m.}/Y_{beam}~=~Y^{red}$ at different energies. For this, in Fig.\ref{fig:1}, we display the 
rapidity distribution for the light charged fragments (LCP's) at different incident energies. The rapidity 
distribution is found to
vary drastically throughout the range from $|Y^{red}|~\le~1.75$. 

This is indicating the 
compressed or participant zone around $0$ value and decay into spectator zone towards both side of the $0$ value. 
However around $0$ value, the region between $-0.1$ to $0.1$ is specified as mid-rapidity region in the 
Literature \cite{Andr055, Zhan065}.  
The decay from $-0.1$ towards negative side approach to target like spectator, while, other side is known as projectile 
like spectator. Interestingly, these regions are found to affect drastically with incident energy. With increase 
in incident energy, number of particles are increasing in the region from $-1.25$ to $1.25$, 
while decreasing away from this region on both side. The inset in the figure is showing the clear view of change in 
behavior around $-1.25$, which is also true for other side also. The rapidity distribution is also used as thermalization source in the literature many 
times \cite{Kuma105}. On the basis of this, we have divided the rapidity distribution in five different zones out of
which three are with including $|Y^{red}|~<~0.1$(participant zone), while other two are with 
excluding this region (target and projectile spectator).They are as:
(i) From -0.1 $\le$ $Y^{red}$ $\le$ 0.1 with increment of 0.2 on both side upto  
-1.5 $\le$ $Y^{red}$ $\le$ 1.5. (ii)   From 0 $\le$ $Y^{red}$ $\le$ 0.1 with increment of 0.2 on latter 
side upto  
0 $\le$ $Y^{red}$ $\le$ 1.5. (iii) From -0.1 $\le$ $Y^{red}$ $\le$ 0 with decrement of 0.2 the on former 
side upto  
-1.5 $\le$ $Y^{red}$ $\le$ 0. (iv) From $Y^{red}$ $\ge$ 0.1 with increment of 0.2 upto $Y^{red}$ $\ge$ 1.5. and 
(v)From $Y^{red}$ $<$ -0.1 with decrement of 0.2 upto $Y^{red}$ $<$ -1.5.\\
%%%%%%%%%%%%%%%%%%%%%%%%%%%%%%%%%%%%%%%%%%%%%%%%%%%%%%%%%%%%%%%%%%%%%%%%%%%%%%%%%%%%%%%%%%%%%%%%%%%%%
\begin{figure}
\includegraphics{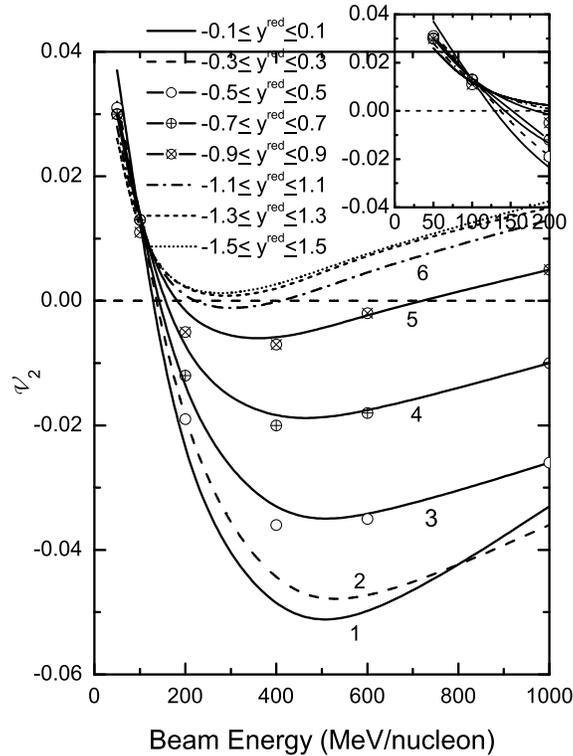}
\caption{\label{fig:2}The incident energy dependence of elliptical flow for LCP's collectively for 
projectile as well as target matter including mid-rapidity region. The different lines are at different size of the 
rapidity bin, which includes the participant as well as spectator matter.}
\end{figure}
%%%%%%%%%%%%%%%%%%%%%%%%%%%%%%%%%%%%%%%%%%%%%%%%%%%%%%%%%%%%%%%%%%%%%%%%%%%%%%%%%%%%%%%%%%%%%%%%%%%%%%%%%%%%%%%%5
Let us now understand the effect of these rapidity cuts on the excitation function of elliptical flow.
In Fig.\ref{fig:2}, we present the excitation function of elliptical flow for the first condition. 
This condition is the mixture of the participant as well as spectator zone from projectile as well as target matter. 
There are two points to discuss here. One is the change in the elliptical flow with incident energy and other is to see the effect of bin size on elliptical flow. 
The elliptical flow evolves from a positive value, rotational-like, emission to an negative value, collective 
expansion, with increase in the incident energy. In other words, transition from the in-plane to out-of-plane takes place.
The energy at which this transition takes place is known as transition energy. This transition is due to 
the competition between the mean field at low incident energy and NN collisions at high incident energies. The incident 
energy dependence of directed flow also show such type of transitions from negative to positive value, which is 
know as balance energy \cite{Sood065}. 
After this transition, the strength of collective expansion overcomes the rotational like motion \cite{Wang96}.
This leads to increase of out-of-plane emission towards a maximum around 400 MeV/nucleon. 
This maxima is further supported by the nuclear stopping at 400 MeV/nucleon \cite{Reis04}.

Beyond this energy, 
elliptical flow decreases indicating a transition to in-plane preferentially emission \cite{Pink995}. This rise and fall 
behavior of the elliptical flow in the expansion region is due to the variation in the passing time $t_{pass}$ of the 
spectator and expansion time of the participant zone \cite{Andr055}. In a simple participant spectator model, 
$t_{pass}~=~2R/(\gamma_{s}v_s)$, where R is the radius of the nucleus at rest, $v_s$ is the spectator velocity in
c.m. and $\gamma_s$ the corresponding Lorentz factor. Due to the comparable value of the passing time and 
expansion time in the collective expansion region up to 400 MeV/nucleon, the elliptical flow results an 
interplay between fireball expansion and spectator shadowing. In other words, due to the comparable size of two times(the fireball expansion and spectator shadowing), 
the participant particles which will come in the way of spectator shadowing are deflected towards the out-of-plane and 
hence more squeeze out. However, in the energy range from 400 MeV/nucleon to 
1.49 GeV/nucleon, $t_{pass}$ decreases from 30 to 16 fm/c (not shown here), implying that overall the expansion gets about two times
faster in this energy region \cite{Andr055}. This is supported by the average expansion velocities extracted from the 
particle spectra \cite{Wang96}. In this case, corresponding to much shorter passing time compared to the expansion time, 
the participant zone is affected by the shadowing of the spectator at very early times, however no shadowing
effect is observed at later time where expansion of the participant matter is still happening. In other words,
the participant particles at later times are not blocked by the shadowing of spectator and hence decrease
is observed in the out-of-plane emission after 400 MeV/nucleon.\\
%%%%%%%%%%%%%%%%%%%%%%%%%%%%%%%%%%%%%%%%%%%%%%%%%%%%%%%%%%%%%%%%%%%%%%%%%%%%%%%%%%%%%%%%%%%%%%%%%%%
On the other hand, with increase in the rapidity region, the transition energy is found to affect to a great extent. 
In literature, the balance energy using directed flow is also calculated, but was over the entire rapidity 
Distribution \cite{Sood065}. 
The rapidity distribution affects many phenomena in intermediate energy region like particle 
production, nuclear stopping \cite{Kuma105} and now the elliptical flow. The transition energy 
increases with rapidity region between $|Y^{red}|~\le~0.1$ and $|Y^{red}|~\le~1.5$. 
With the increase in the rapidity region, 
dominance of the spectator matter from projectile as well as target takes place that will further result in the dominance of 
the mean field up to higher energies. After the transition energy, the collective expansion is  found to have less squeeze
out with an increase in the rapidity region. In this region, the spectator zone contribution along with 
participant zone starts to come in play. As we know, the passing time for the spectator is very less compared to 
expansion time of the participant zone, leading to the decreasing effect of the spectator shadowing on the
participant zone. The chances of the participant to move in-plane increases with increase in the rapidity bin and 
hence less squeeze out is observed with increase in the size of the rapidity bin. If one see carefully, no 
transition is observed after $|Y^{red}|~\le~1.1$. This is due to negligible effect of the participant zone 
compared to the spectator zone. The inset in the figure shows interesting results: the intersection 
of all the rapidity bins takes place at a particular incident energy.  Below and above this energy, incident energy
dependence of elliptical flow is changing the behavior with rapidity distribution. One can have a good 
study of elliptical flow below and above this particular incident energy with variation in the rapidity distribution.\\
\begin{figure}
\includegraphics{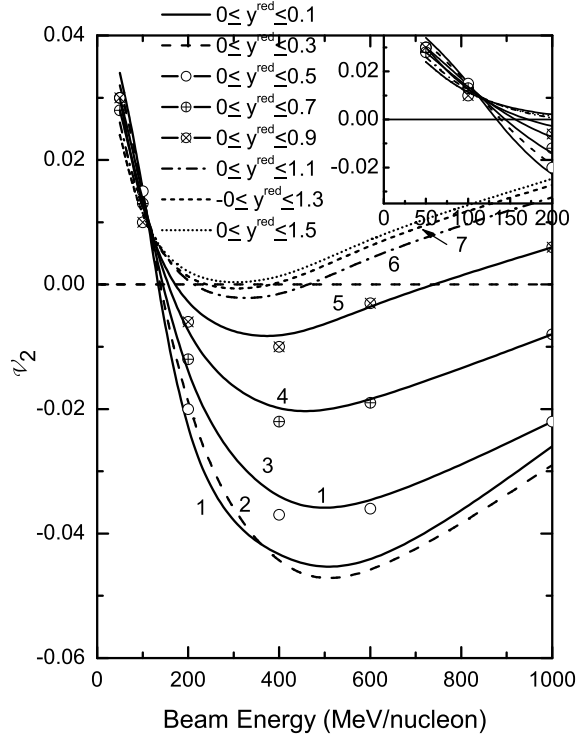}
\caption{\label{fig:3} Same as in fig.\ref{fig:2}, but the contribution for fragments is from the 
 projectile like matter only. All the bins includes mid-rapidity region.}
\end{figure}

\begin{figure}
\includegraphics{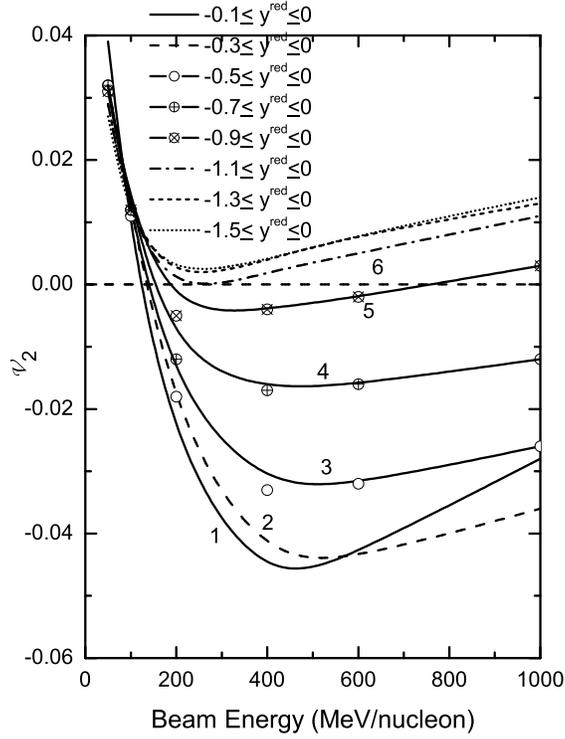}
\caption{\label{fig:4}Same as in fig.\ref{fig:2}, but the contribution for fragments is from the 
 target like matter only. All the bins includes mid-rapidity region. }
\end{figure}
%%%%%%%%%%%%%%%%%%%%%%%%%%%%%%%%%%%%%%%%%%%%%%%%%%%%%%%%%%%%%%%%%%%%%%%%%%%%%%%%%%%%%%%%%%%%%%%%%%%%%%%%%%%%%%%%
In Fig.\ref{fig:2}, we had displayed the effect on the participant as well as spectator matter from the projectile as
well as target collectively including the distribution at $|Y^{red}|~<~0.1$. It becomes quite interesting to 
study the participant and spectator matter contribution on the incident energy dependence of elliptical flow
for projectile as well as target matter separately. For this, in Figs.\ref{fig:3} and \ref{fig:4}, the incident energy
dependence of the elliptical flow are displayed from the mid-rapidity to spectator zone for projectile and target matter.
Both of the figures are following the universal behavior of the in-plane to out-of-plane 
emission with increase in incident energy, as is displayed in Fig.\ref{fig:2}. On the other hand, 
the effect of rapidity bins is also quite similar as seen in of Fig.\ref{fig:2}. 

If one observes the maximum squeeze out values in Figs. \ref{fig:2}-\ref{fig:4} around
400 MeV/nucleon, it is found that less squeeze out is observed for projectile matter and spectator matter, separately
as compared to projectile and target matter, collectively. However, it is almost same for the projectile or 
target matter, separately. This is obvious as when rapidity bin is from 
0 to 0.1 or -0.1 to 0, then participant zone is less compared to the region -0.1 to 0.1. This can 
be further clarify from the Fig.\ref{fig:1}, where rapidity distribution for different bins is 
displayed. This is true for 
all the bins under investigation. The effect of rapidity distribution on the transition energy will be 
discussed later under different conditions.\\
%%%%%%%%%%%%%%%%%%%%%%%%%%%%%%%%%%%%%%%%%%%%%%%%%%%%%%%%%%%%%%%%%%%%%%%%%%%%%%%%%5
\begin{figure}
\includegraphics{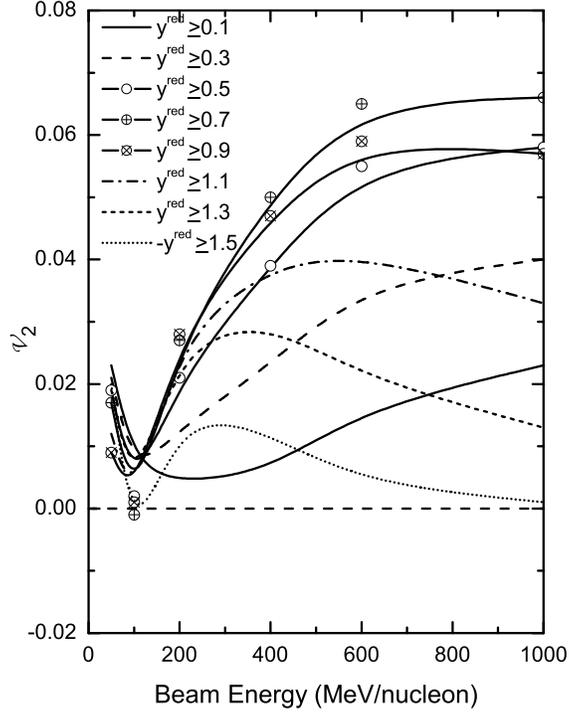}
\caption{\label{fig:5}Same as in fig.\ref{fig:2}, but the contribution for fragments is from the 
 projectile like matter only. All the bins excludes the mid-rapidity region. In this case, we move from 
participant+spectator contribution towards spectator contribution.}
\end{figure}
\begin{figure}
\includegraphics{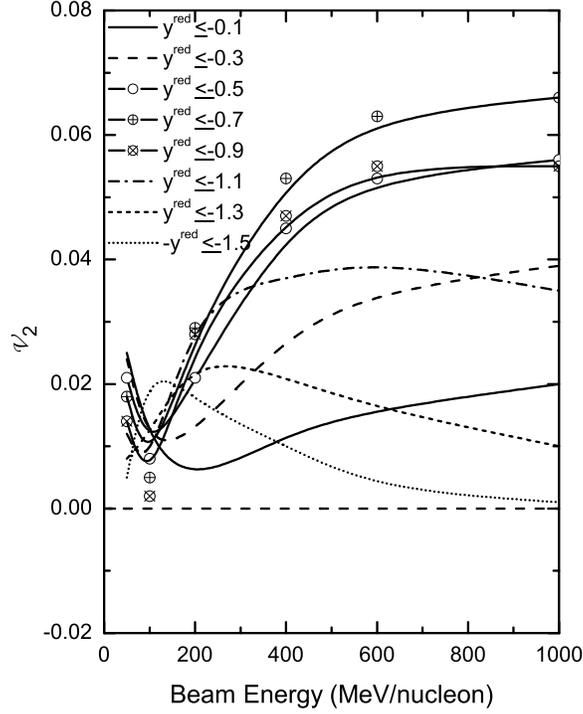}
\caption{\label{fig:6} Same as in fig.\ref{fig:2}, but the contribution for fragments is from the 
 target like matter only. All the bins excludes the mid-rapidity region. In this case, we move from 
participant+spectator contribution towards spectator contribution.}
\end{figure}

One notices from the above figures that major contribution for elliptical flow comes from the mid rapidity
region. One is further interested to know the fall of elliptical flow if the mid-rapidity region is excluded.
For this, by excluding  the $|Y^{red}|~<~0.1$ region, we have displayed the incident energy dependence 
for projectile and target matter from participant to purely projectile or target spectator matter in 
Fig.\ref{fig:5} and Fig.\ref{fig:6}, 
respectively. The noted behavior is entirely different  
compared to previous three figures. The differences are:
(a) The behavior of rise and fall in the elliptical flow at a particular incident energy is obtained as were in the 
previous figures, but, no transition is observed from in-plane to out-of-plane emission. Over all the 
incident energies, the value of elliptical flow remains positive, however, some competition is 
observed at low incident energies around 150 MeV/nucleon. This is due to the dominance of attractive mean field 
from the spectator matter, which restricts the effects of participant zone.\\     
(b) A higher value of in-plane flow is observed at high incident energies compared to low incident energies 
upto $Y^{red}~\ge~0.7$ in Fig.\ref{fig:5} for projectile matter and $Y^{red}~\le~-0.7$ for target matter, which
is in contrast to the previous figures. This is due to the effect that at low incident energies, the passing time of 
spectator and expansion time of the participant and comparable, while, with increase in the incident energy, the 
passing time is found to be decrease as compared to expansion time. This will reduce the shadowing effect at 
higher incident energies on the particles of the participant zone, results in the enhanced in plane flow. Moreover,
more higher value of in-plane flow at higher incident energies as compared to lower one is due to the absence of most
compressible region from the region $|Y^{red}|~\le~-0.1$, which competes with the mean field of the spectator to 
a great extent.  \\ 
(c) With increase in the rapidity region from participant and spectator ($Y^{red}~\ge~0.1$ )to purely spectator matter 
($Y^{red}~\ge~1.5$ ), the 
dominance of in-plane flow 
takes place upto $Y^{red}~\ge~0.7$ and after that suddenly fall in the in-plane flow takes place upto $Y^{red}~\ge~1.5$ in
Fig.\ref{fig:5}, which is also true for the Fig.\ref{fig:6}. The fall in the in-plane flow after $Y^{red}~\ge~0.7$ 
can be explained with the help of the Fig.\ref{fig:1}. After this region, the contribution of the  
participant is almost negligible and size of the spectator also decreases. Due to the decreasing size of the spectator matter, the violence of the incident 
energy as well as short passing time of the spectator, they start to fly out-of-plane, which were enjoying in-plane 
earlier due to their heavier size. Hence, minimum in-plane flow is observed for $Y^{red}~\ge~1.5$ where only 
5-7 particles exist as is clear from the Fig.\ref{fig:1}. The same behavior is observed in the Fig.\ref{fig:6}.\\
\begin{figure*}
\includegraphics{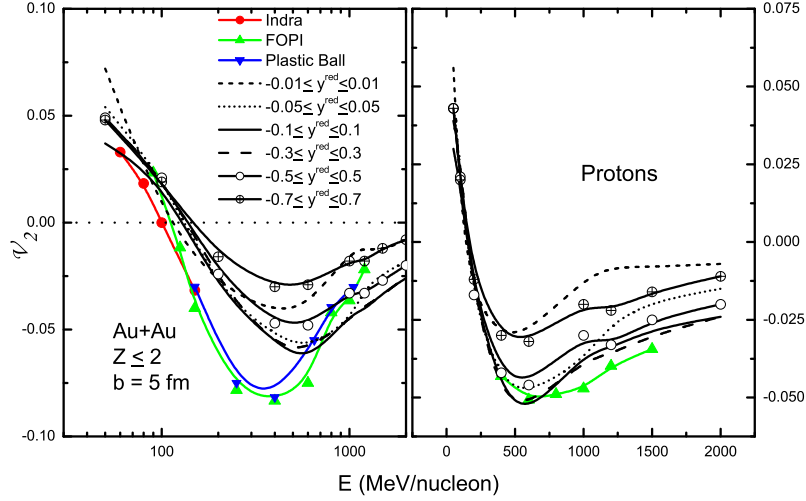}
\caption{\label{fig:7} (Color online) Comparison of the results of incident energy dependence of elliptical flow 
including different rapidity bins with the experimental results of different collaborations \cite{Luka055, Andr055}.
The left hand side is for Z $\le$ 2 particles, while, right hand side is for the protons.}
\end{figure*}
\begin{figure}
\includegraphics{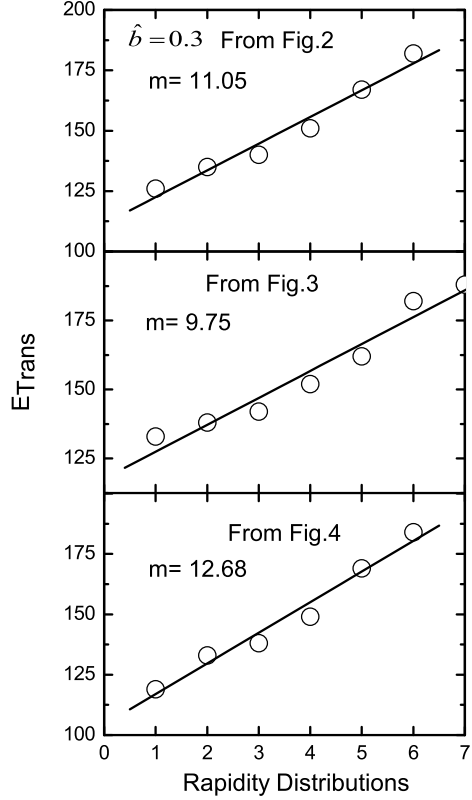}
\caption{\label{fig:8}The dependence of rapidity distributions on the transition energy. The transition energies 
are extracted from Fig.\ref{fig:2}, \ref{fig:3}and \ref{fig:4} for top, middle and bottom panels,
respectively. The numbering from 1to 7 is representing the different rapidity bin from the respective figures
mentioned above. All the panels are parameterized with the straight line interpolation $Y = mX~+~C$, where m is the 
slope, displayed in respective panels.}
\end{figure}
%%%%%%%%%%%%%%%%%%%%%%%%%%%%%%%%%%%%%%%%%%%%%%%%%%%%%%%%%%%%%%%%%%%%%%%%%%%%%%%%%%%%%%%%%%%%%%%%%%%%%%%%%%%%%%%%%%%%5

Going through all the aspects of rapidity distribution, it is important to include the mid-rapidity region 
$|Y^{red}|~<~0.1$ in the study of elliptical flow in heavy-ion collisions. In order to compare the findings with the experimental
one, one must have the transition from in-plane to out-of-plane, which are observed in Figs.\ref{fig:2}-\ref{fig:4}. Moreover, more negative values are observed in Fig.\ref{fig:2} compared to other two one. 
From this discussion, it is fruitful to compare the findings of Fig.\ref{fig:2} with experimental findings of 
different collaborations. This is displayed in Fig.\ref{fig:7} for LCP's (Z~$\le$~2) and protons. The 
curves have the similar trend as displayed in Fig.\ref{fig:2}. Interestingly, it is observed that there is a competition
in the rapidity bin $|Y^{red}|\le 0.1$ and $|Y^{red}|\le 0.3$. After that systematic deviation is observed in 
the elliptical flow values from the data values with increase in the rapidity bin. There are some deviation in the
middle region in case of LCP's between theory and data, while data is well explained for the protons. From here,
it is concluded that one can vary the mid rapidity region from the $|Y^{red}|\le 0.1$ to $|Y^{red}|\le 0.3$ to 
get the better agreement with the experimental one. However, the discrepancy with the data for
 LCP's can be reduced by varying the isospin-dependent cross sections.\\
%%%%%%%%%%%%%%%%%%%%%%%%%%%%%%%%%%%%%%%%%%%%%%%%%%%%%%%%%%%%%%%%%%%%%%%%%%%%%%%%%%%%%%%%%%%%%%%%%%%5
Last, but not least, the rapidity distribution dependence of the transition energy is displayed in the Fig.\ref{fig:8}.
The top, middle and bottom panels are representing the transition energies extracted  
from Fig.\ref{fig:2}-\ref{fig:4}, respectively. The numbering from 1 to 7 in the figure is representing the bin size from the respective figures. All the curves are fitted with straight line equation $Y~=~mX~+~C$, where m is slope of line and C is a constant. The transition energy is found to be sensitive
towards the different bins of rapidity distributions. It is observed that transition energy is found to increase
with the size of the rapidity bin for light charged particles. 
The necessary condition for the transition energy is that the mid-rapidity region must
be included in the rapidity distribution bin. It is also observed that no transition energy is obtained when the
rapidity distribution region extends away from $|Y_{red}|~\le~1.1$. This is due to the dominance of the spectator matter, which enjoys in-plane as compared to out-of-plane. The transition energy is found to be weakly sensitive
towards the choice of the different type of rapidity distributions (projectile-target or projectile or target) while we have included the  
mid-rapidity region. However, transition energy is found to disappear when mid-rapidity region is excluded. 
In other studies, system size dependence of the transition energy is also studied by us and others for different kind of 
Fragments \cite{Zhan065}. Recently, we have studied the fragment size dependence of transition energy 
under the influence of 
different equations of state, nucleon-nucleon cross sections etc \cite{Sood065}. These studies were made at a fixed rapidity bin. The detailed analysis of the transition energy with rapidity distributions has revealed many interesting aspects for the first time.\\

For more interest, in the near future, we are performing the comparative study of different QMD and IQMD simulations. From the preliminary results, it is observed that isospin content of the colliding partners $(N/Z~=~1.49)$ is supposed to playing the appreciable role in the analysis of elliptical flow in intermediate energy heavy-ion collisions in term of (i) closeness of the results with experimental findings of different collaborations and (ii) effect on the elliptical flow with change in the rapidity bins. The detailed analysis of these findings will be presented in the near future.\\ 
   
\section{Conclusion} 
Within the semi classical transport simulations of energetic semi central collisions of 
$_{79}Au^{197}~+~_{79}Au^{197}$ reaction, we have carried out a new investigation of the interplay between the 
participant and
spectator regions in term of rapidity distributions. 
The maxima and minima in the incident energy dependence of elliptical flow is produced due to the 
different contributions of passing time of the spectator and expansion time of the participant. The shadowing of 
the spectator matter plays important role up to later times due to the comparable magnitude of 
the passing and expansion time up to energy 400 MeV/nucleon, however at high energies, shadowing 
effect is dominant only at earlier times due to the shorter passing time as compared to expansion time. 
 The transition from in-plane to out-of-plane is observed only when the mid-rapidity region is included in the
rapidity bin, otherwise, no transition is observed. The transition energy is found to be strongly dependent on 
the size of the rapidity bin, while, weakly dependent on the type of the rapidity distributions. The transition
energy is parameterized with a straight line interpolation. Comparison with experimental bin, reveals  
the competition is observed between the rapidity bin of 
$|Y^{red}|~\le~0.1$ and $|Y^{red}|~\le~ 0.3$. To remove this discrepancy in the middle region 
for LCP's, one has to reduce the strength of nucleon-nucleon cross-section.\\

%%%%%%%%%%%%%%%%%%%%%%%%%%%%%%%%%%%%%%%%%%%%%%%%%%%%%%%%%%%%%%%%%%%%%%%%%%%%%%%%%%%%%%%%%%%%%%%%%%%%%%%

\begin{acknowledgments}
This work has been supported by the Grant no. 03(1062)06/ EMR-II, from the Council of Scientific and
Industrial Research (CSIR) New Delhi, Govt. of India.\\
\end{acknowledgments}
%\newpage %Just because of unusual number of tables stacked at end

\end{document}